\documentclass[prl,twocolumn,showpacs]{revtex4-1}

\usepackage{graphicx}
\usepackage{ulem}
\usepackage{dcolumn}
\usepackage{bm}
\usepackage{natbib}

\begin{document}

\title{Spherical Shell Cosmological Model and Uniformity of Cosmic Microwave Background Radiation}

\author{Branislav Vlahovic}
\email{vlahovic@nccu.edu}
\affiliation{Department  of Physics, North Carolina Central University, 1801 Fayetteville 	Street, Durham, NC 27707 USA.}


\begin{abstract}
\noindent

Considered is spherical shell as a model for visible universe and parameters that such model must have to comply with the observable data. The topology of the model requires that motion of all galaxies and light must be confined inside a spherical shell. Consequently the observable universe cannot be defined as a sphere centered on the observer, rather it is an arc length within the volume of the spherical shell. The radius of the shell is 4.46 $\pm$ 0.06 Gpc, which is for factor $\pi$ smaller than radius of a corresponding 3-sphere. However the event horizon, defined as the arc length inside the shell, has the size of 14.0 $\pm$ 0.2 Gpc, which is in agreement with the observable data. The model predicts, without inflation theory, the isotropy and uniformity of the CMB. It predicts the correct value for the Hubble constant $H_0$ = 67.26 $\pm$ 0.90 km/s/Mpc, the cosmic expansion rate $H(z)$, and the speed of the event horizon in agreement with observations. The theoretical suport for shell model comes from general relativity, curvature of space by mass, and from holographic principle. The model explains the reason for the established discrepancy between the non-covariant version of the holographic principle and the calculated dimensionless entropy $(S/k)$ for the visible universe, which exceeds the entropy of a black hole.   The model is in accordance with the distribution of radio sources in space, type Ia data, and data from the Hubble Ultra Deep Field optical and near-infrared survey.
\end{abstract}

\maketitle

\section{Introduction}

The paradigm of $\Lambda$CDM cosmology works impressively well and with concept of inflation it explains universe after the time of decoupling. General relativity and standard model can predict with high accuracy decrease in orbital period of binary pulsar and angular power spectrum of CMB. Experimental data from BAO, CMB, SN Ia, and observations of large scale structures allow to put some constrains to cosmological models and parameters.

 However there are still a few concerns. After all efforts there is no detection of dark matter and there are significant problems in theoretical description of dark energy. For that reason there are numerous attempts for alternative cosmologies that could modify GR theories (modify  gravity $G_{\mu\nu}$, e.g. brane-worlds, modified action theories, f(R) gravity, tensor gravity, higher dimensional gravity)
 \begin{equation} \label{GR__1}
 G_{\mu\nu} = 8\pi G\tilde{T}_{\mu\nu} \hspace{0.1in} where: \hspace{0.1in} \tilde{T}_{\mu\nu} \equiv T_{\mu\nu} - \frac {\Lambda} {8{\pi}G}g_{\mu\nu}
\end{equation}
 or modify matter theories (modify $\tilde{T}_{\mu\nu}$, inhomogeneous universe, new matter, new interactions, quintessence, Chaplyging gas, k-essence).

The modification to GR will led to modified Friedmann and acceleration equations, e.g. modification of f(r) gravity will produce extra first four terms on the left side of equation:
\begin{equation} \label{GR__2}
 -H^2 fr+ \frac {a^2} {6} f + \frac {3} {2} H \mbox{\.f} r + \frac {1} {2} \mbox{\"f} r + \frac {\mbox{\"a}} {a} = - \frac {4\pi G} {3}(\rho + 3P)
\end{equation}
and for instance higher dimensional gravity will add the first term on the left:
\begin{equation} \label{GR__3}
  \frac {-\mbox{\.H}} {r_c} + \frac {\mbox{\"a}} {a} = - \frac {4\pi G} {3}(\rho + 3P).
\end{equation}
Because current data are not enough to discriminate between the GR and alternative theories there are ongoing and planned surveys to provide more accurate data and additional tests, such as: LSS observations (galaxy positions, weak lensing, redshift space distortions),  dark energy survey, Euclid (an ESA mission to map the geometry of the dark Universe ), evolutionary map of the universe, Westerbork observation of the deep aperitif northern sky.
The goal of this paper is to consider a new cosmological model that will not require alternative GR and gravity or matter modification, but will allow to describe observable data without inflation.

It is well known that in the Big-Bang models homogeneity of space cannot be explained, it is simply assumed in initial conditions. The curvature and physical properties of the regions of the space which have never been in causal contact  and should not be correlated are taken to be indistinguishable. In CMB spectrum points further apart than $2^0$ should be not correlated, but correlations up to $\sim 60^0$ are observed. Homogeneity on the level of $10^{-5}$ is explained by inflation era. However, arguments in favor of inflation only exist if space was already homogeneous before inflation. If the pre inflationary universe was not already homogenous inflation will not lead to homogeneity \cite{Goldwirth}. So, the homogeneity problem is pushed only back in time, because Big-Bang itself is taken to be inherently free of correlations. In addition, no satisfactory model for inflation exist. We will show that proposed cosmological model could explain uniformity of CMB without inflation theory.

GR does not specify the topology of the space, Einstein's equation describe only local property of the spacetime, but do not fix the global structure, topology of spacetime.  For the same matrix element different topologies can correspond leaving the possibility for a new universe models. Universe is correctly described by Friedmann equations but the values of the cosmic parameters are not known accurately to determine topology and curvature. We do not know the real value of the comoving space curvature radius or the present value of the scale factor $R(t_o)=R_o$, the only cosmological length that is directly observable is the Hubble length $L_{Hub}=cH_o^{-1}$. For non flat universe $R_o=L_{Hub}/\sqrt{|\Omega + \Lambda -1|}$ and there is no scale for the flat case.

\section{Spherical Shell Model}

We will consider a model in which the universe is an expanding spherical shell with thickness much smaller than its radius. Such model can be justified by the short time interval of the Big-Bang universe creation and since then continuous Hubble expansion. We may assume, as it has been always emphasized, that galaxies do not move through space and that the universe is not expanding into empty space around it, for space does not exist apart from the universe.

\begin{figure}
		\includegraphics[width=8cm]{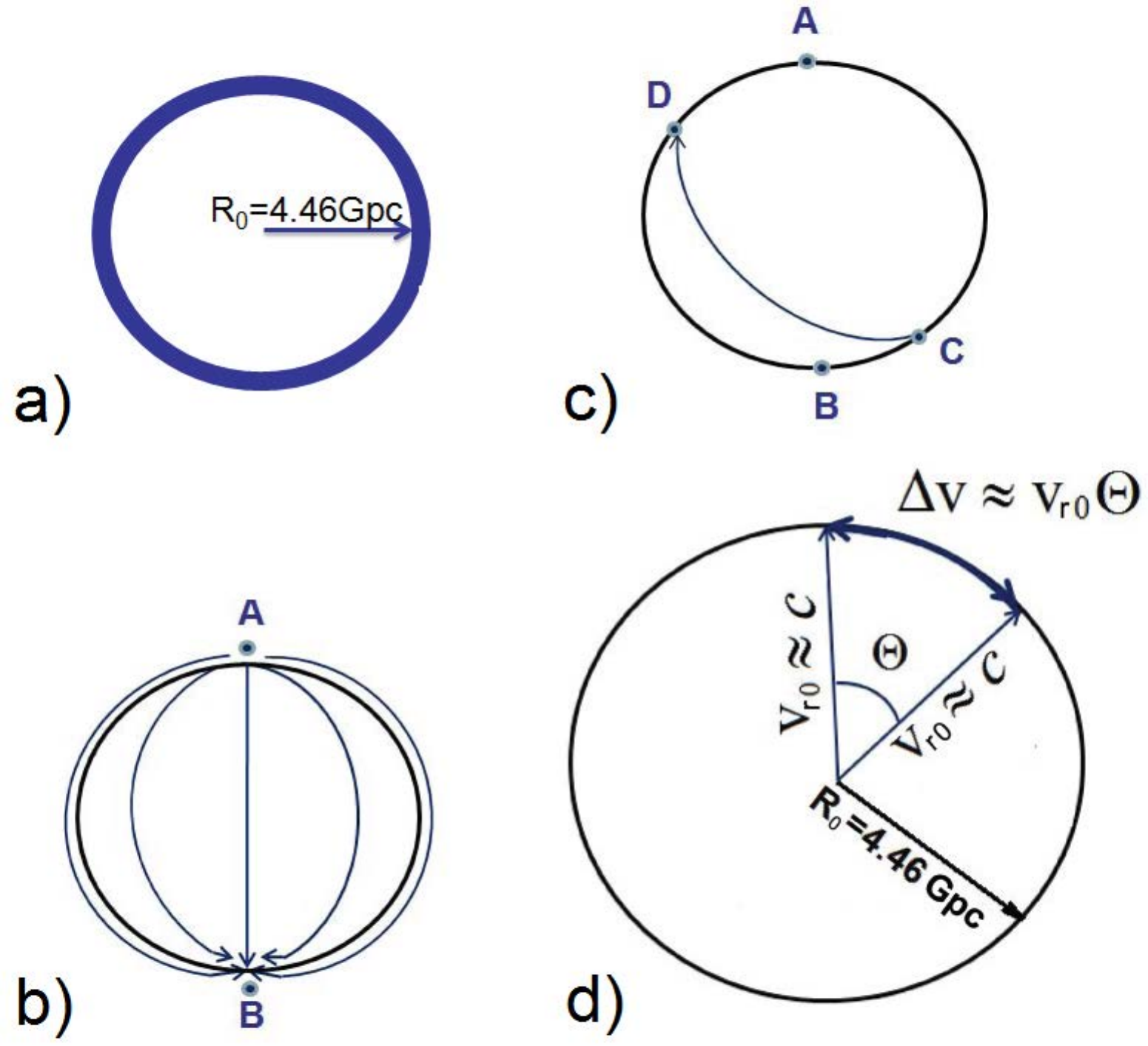}
		\caption{\label{fig2}
 a) The visible universe as an expanding shell with thickness much smaller than radius, b) Observable universe, as seen by an observer from the point A, is a volume of the shell, with event horizon located in the point B, c) CMB visible from Earth (by observer in point A) is originated in the antipodal point B and CMB visible from another place in the universe (point C) is emitted in the point D, d) The visible universe as a surface of the sphere with radius $R_0$ = 4.46 Gpc that expands with radial speed close to the speed of light.
}
\end{figure}

There is no other space than that associated with the spherical shell. The motions of all galaxies and propagation of the light are confined to the volume of the shell, which expands with a radial velocity. As we will see this will have significant implication on the interpretation of the data.  The light must follow geodesic lines, so like in torus models it is not traveling straight, it is bend. This will define observable universe for our model as the largest visible volume (from the point of the observer) inside the spherical shell that represents the universe, Fig. \ref{fig2}a). By this definition the cosmological horizon distance will be the largest possible distance inside the spherical shell. If for instance the universe is the same size as the observable universe and an observer is located at point A, the particle horizon for that observer will be the point B at the antipode and the observable universe will be the entire spherical shell. This is shown in Fig. \ref{fig2}b.

The dynamics of shell models has been investigated earlier. It was first introduced by Israel \cite{Israel}, in the framework of the special-relativity by \cite{Czachor}, and a systematic study in the framework of general relativity is done for instance in  \cite{Berezin} and \cite{Krisch}.  However, our focus will be very different. We will present significant implications of the shell model when combined with a new proposed interpretation of experimental data.

The spherical shell model is in agreement with the main cosmological principles, isotropy and homogeneity of the space and as it will be shown it also satisfies Friedman-Lema\^itre model. The model also must be in agreement with the observational data, the age of the universe and direct observations of matter and density. The dimension size of the shell (the arc length from pole to  antipode) must correspond to the present size of the cosmological horizon, and the thickness of the shell must have the minimum size to explain present observation constrains: ghost images of sources; distribution and periodicity of clusters, super clusters, quasars, and gamma-ray bursts; statistical analysis of reciprocal distances between celestial objects; and other limits obtained from CMB (uniformity and weak angular fluctuations of the CMB). As for example, from the statistical analysis of Abell catalog of spatial separation of clusters it appears that the shell thickness should be at least about 1 Gpc  \cite{Lehoucq}.

In the current $\Lambda$CDM model the visible universe is defined as as a sphere centered on the observer and from our perspective it appears that the radius is $R_0$ = 14.0 $\pm$ 0.2 Gpc (about 45.7 Gly).  The value $R_0$ is the particle horizon and the quoted result corresponds to the direct WMAP7 measurements and the recombination redshift $z$ = 1090 $\pm$ 1 \cite{2}. In the standard FLRW model
  \begin{equation} \label{radius__1}
 R_o = a(t) \int_{t'=0}^{t}\frac{c}{a(t')}dt',
 \end{equation}
where    \begin{equation} \label{radius__2}
\frac{da}{dt} = \sqrt{\frac{\Omega_r}{a^2} + \frac{\Omega_m}{a} + \frac{\Omega_\Lambda}{a^{-2}}}.
\end{equation}
The  $R_0$ = 14.0 $\pm$ 0.2 Gp corresponds to the following combinations of the parameters: $\Omega_m h^2 =0.136 \pm 0.003$, $\Omega_r = \frac{8\pi G}{3H^2} \frac{\pi^2 k^4 T^4}{15 c^5\hbar^4}$, $\Omega_{\Lambda} = 1 - \Omega_r - \Omega_m$.
In our shell model we must obtain the same value for the particle horizon, which is now not the radius of the sphere, but an arc distance from pole to antipode. Therefore, the curvature radius of our spherical shell will be for factor $\pi$ smaller than in the standard FLRW model. This will result in significantly different density, which will be at least for factor of $\pi^3$ (depending on the thickness of the shell) bigger in the spherical shell model.

\section{Important implication of the model, uniformity of CMB}

In addition to the significantly higher prediction for density, the most important implication of the shell model  is prediction of the uniformity of the CMB without inflation.
In the proposed spherical shell model, looking from the position of our galaxy (marked by A), the place of decoupling (the surface of last scattering, the source of CMB) is at the antipode (marked by B), Fig. \ref{fig2}c. Regardless of the direction we chose to measure CMB (for instance from point A looking in any arbitrary chosen direction), we will always measure CMB at the antipodal point (in this example point B). The reason for this is that the length of the arc on the sphere represents distance, which also represents past time. When the visible universe is of the same size as the whole universe, the point that is at the largest distance away from the point A (point B) represents the surface of last scatter, since we cannot see beyond that distance.

It is important to notice that measuring the same CMB by looking in the opposite directions of the universe does not represent or reflect the uniformity of the universe at the time of decoupling, because we always measure CMB originated from the same point regardless of the direction of observation.  For that reason we always must obtain the same result. If from point A we observe to the right, left, backward or forward we will always measure CMB originated from the point B. Small variations for the CMB are possible and they are observed, but they are the result of the interaction between matter and light during its travel. For instance, depending on the direction we choose to measure CMB, light will travel from point B to A through different galaxies and will interact with different amounts of matter, which will result in the small observed variations of CMB. The observed fluctuations in the CMB are therefore created as the photons pass through nearby large scale structures by the integrated Sachs-Wolfe effect. The correlation between the fluctuations in the CMB and the matter distribution is well established \cite{4a}\cite{4b}\cite{4c}\cite{4d}.

To establish a connection between the uniformity of the earlier universe at the time of decoupling and the CMB we will need to make a completely different kind of measurements of the CMB. We can see the CMB in any direction we can look in the sky. However, we must keep in mind that the CMB emitted by the matter that would ultimately form for instance the Milky Way is long gone. It left our part of the universe at the speed of light billions of years ago and now forms the CMB for observers in remote parts of the universe, actually exactly for an observer at the antipodal point B. For instance, if we perform measurement of the CMB at the point C, we will measure the CMB emitted by matter at the point D, Fig. \ref{fig2}c.  To measure uniformity of the universe at the time of decoupling we will need to measure the CMB in at least two different points on the shell. If, for instance, the measurements from points A (CMB originated in B) and C (CMB originated at D) give the same result, then and only then may we speak about the uniformity of the CMB and uniformity of the universe at the time of decoupling. However, such measurements are not possible at the present time.

The strongest objection to this interpretation of the uniformity of CMB could be that observer located at the pole cannot see CMB originated at antipodal point, because the light originated at point B does not have enough time to reach observer at point A. In a matter dominated universe, light from an antipodal point  on  FLRW expanding sphere, can never reach an observer during the expanding phase $H > 0$. For closed universe in FRW  metric
\begin{equation} \label{metric}
                      ds^2= c^2dt^2-a^2(t)[d\chi^2 + sin^2\chi(d\theta^2 + Sin^2\theta d\phi^2)],
\end{equation}
where $\chi=\int cdt/a(t)$, which is by definition conformal time coordinate or arc parameter time $\eta$. Substituting for $ds^2 = 0$ and assuming that observer is located at $\chi=0$ and CMB is originated at antipodal point $\chi_{B} = \pi$, the light will reach observer when $\eta = \pi$
\begin{equation} \label{eta}
                      \eta_0 - \eta = \pi = \int_{0}^{t_0} \frac{cdt}{a(t)} = \frac{c}{a_0}\int_{0}^{z}\frac {dz'}{H(z')}.
\end{equation}
For matter dominated universe $a(\eta)=A(1-cos\eta), ct=A(\eta-sin\eta)$. Therefore when antipodal light reach observer at $\eta = \pi$ the expansion will be at the maximum and  $H=0$. However this is not true for closed universe with a cosmological term.  Because of the $\Lambda$ term universe may not collapse and a light from antipodal point can reach observer during the expansion epoch. Also as we will show later the dynamics of the spherical shell is different from here considered dynamics of 3-sphere.

Another serious objection could be that the angular power spectrum of the CMB temperature fluctuations $\Delta T/T$
predicted by $\Lambda CDM$ model agrees well with observations and that such agreement cannot be obtained for spherical shell model, because power spectrum coefficients $a_{lm}$ are model dependent and will be different for 3-sphere and spherical shell. However, situation is not so simple because it is well known that the same CMB anisotropy spectra can be produced by two different models having different combinations of the parameters \cite{Efstathiou},\cite{Efstathiou1}. More importantly there is no reason to check if ${\Lambda}CDM$ and spherical shell are degenerate models, which could produce the same anisotropy spectra, because the CMB power spectrum does not have any meaning for the spherical shell model.

The CMB temperature fluctuations $\Delta T/T$ are usually written in terms of a multipoles expansion on the celestial sphere:
\begin{equation} \label{deltaT}
                      \frac {{\delta}T} {T} (\theta,\phi) = \sum_{l=2}^{\infty} \sum_{m=-l}^{l}a_{lm} Y_l^m(\theta,\phi).
\end{equation}
However, what is actually directly measured by observations is the angular correlation of the temperature anisotropy
$\langle \frac {\delta T} {T} (\hat{n}_1) \frac {\delta T} {T} (\hat{n}_2) \rangle$ where $cos\theta = \hat{n}_1 \cdot \hat{n}_2$. This is expressed through power spectrum $C_l \equiv \langle |a_{lm}|^2 \rangle$, Legandre polinomials, and the filter function $W_l$ as
   \begin{equation} \label{Ctheta}
                      C(\theta) = \frac {1} {4\pi} \sum_{l} \left[ \frac {l + \frac {1} {2}} {l(l+1)} \right] C_lP_l(cos\theta) W_l.
\end{equation}
        The main contribution to $C_l$ for $l>60$ is from oscillations in the photon-baryon plasma before decoupling. However, in the spherical shell model we cannot see the imprint of these oscillations in the CMB a the time of last scattering, because we are always measuring CMB coming just from the single point, the antipodal point, of the surface of the last scatter. We cannot see the $C_l$ contributions that are form the remaining part of the surface of the last scattering. For instance as we already mentioned, fluctuations from the surface of the last scattering that at the present corresponds to the Milky Way galaxy already left us bilions of years ago. Therefore, in principle, in shell model we cannot obtain the angular correlation of the CMB temperature anisotropy.

        The contribution to $C_l$ for low multipoles $l \leq 60$ is mainly from Sachs-Wolfe effect that relates temperature fluctuations to the integral of variations of the metric evaluated along the line of sight
            \begin{equation} \label{deltaT1}
                      \frac {\delta T} {T} = - \frac {1}{2} \int_{\eta_{rec}}^{\eta_0} \frac {\partial h_{\alpha\beta}} {\partial \eta} e^\alpha e^\beta d\eta.
\end{equation}
  One can argue that the line of sight is similar in 3-sphere and spherical shell model. For instance assume that we are on the surface of a 3-sphere and that propagation of the light is confined to its surface, then the observed distribution of the galaxies on  surface of the 3-sphere and in sphere shell will be the same. Therefore for the spherical shell model we should obtain very similar spectrum for low multipoles $C_l$  as in the standard ${\Lambda}CDM$ model. However, that is only partially true, because curvature radius is different in these two models for factor $\pi$ and density is different at least for factor $\pi^3$. However, if we assume that gravitational potential does not evolve with time, then equation (\ref{deltaT1}) simplifies to
   \begin{equation} \label{deltaTs}
                       \frac {\delta T} {T}    \simeq \frac {1} {3} \frac {\delta \phi} {c^2}.
                       \end{equation}
   Therefore the temperature asymmetry for $C_l \leq 60$ should be similar in both models and one can use this as the test of the spherical shell model. If one obtains with shell model parameters for the low multipoles part of the asymmetry spectrum as observed, then it may indicate that we are leaving in a shell model universe.

\section{Justification for the Spherical Shell Model}

Dynamics of the shell model was considered in special relativity Newtonian (SRN) approach and in full relativistic approach. The similar results are obtained. In SRN approach the speed of the shell expansions depends on the total energy of the system and the maximum speed $c$ is at the beginning of expansion. However, the system that has sufficient amount of energy can retain speed close to $c$ for long time \cite{Czachor}.
\begin{equation} \label{SR1}
                       E = \frac {M_0c^2} {\sqrt{1-(v/c)^2}}  - k\frac {M_0^2} {r} \frac {\delta \phi} {c^2},
                       \end{equation}
which has solutions
\begin{equation} \label{SR2}
                       \frac {v} {c} = \left( 1- \frac {x^2} {({\frac {Ex} {M_0c^2}} +1)^2}\right)^{1/2}
                       \end{equation}
 and
\begin{equation} \label{SR3}
               x_{max} = \frac {1} {1-\frac {E} {M_0c^2}},
                       \end{equation}
where $x=\frac {r} {R}$.

In GR approach at the beginning of the expansion system has infinite speed. However the total size of the expansion is at the end similar to that calculated by SRN approach. GR equation for spherical shell can be written \cite{Israel} as:
\begin{equation} \label{GR1}
                       \mbox{\"R} \left[ (1+\mbox{\.R}^2)^{1/2} + (1+\mbox{\.R}^2 - \frac {2m} {R})^{1/2} \right] = - \frac {m(1+\mbox{\.R}^2)^{1/2}} {R^2},
                       \end{equation}
where $m$ is the gravitational mass of the shell, equivalent to the total energy $E$ in SRN notation. In the units $x=v/R$ and $\mbox{\.R}=v/c$, the solution can be expressed as:
\begin{equation} \label{GR2}
                       \frac {v} {c} = \left[\left (\frac {E} {M_0c^2}+\frac {1} {2x} \right)^2 -1 \right]^{1/2} and
                       \end{equation}
 \begin{equation} \label{GR3}
                       x_{max} = \frac {1} {2} \frac {1} {1- \frac {E} {M_0c^2}}.
                       \end{equation}
 This is qualitatively the same solution as in SRN case for $x\rightarrow \infty$ and $\frac {E} {M_0c^2} > 1$, while for $\frac {E} {M_0c^2} < 1$ it differs from SRN solution for factor $1/2$, which could be due to a different scaling of the sphere radius applied in these two cases.

As mentioned earlier, propagation of light and galaxies is confined to the spherical shell; for that reason we cannot point to the center of the universe. As seen from our galaxy, all other galaxies are moving away from us (and from each other) with the speed $v = v_{r0}\Theta$ (where $v_{r0}$ is the current radial speed of expansion and $\Theta$ is azimuthal angle, Fig. \ref{fig2}d), which is actually the Hubble law $v = H_0 \times distance$. So the spherical shell model predicts value for the Hubble constant that can be calculated from
\begin{equation} \label{GrindEQ__8}
v = v_{r0}\Theta = H_0R_0\Theta
\end{equation}
by inserting for the radius of the shell $R_0$ = 4.46 $\pm$ 0.06 Gpc (which in shell model corresponds to the particle horizon 14.0 $\pm$ 0.2 Gpc, obtained by direct WMAP7 measurement \cite{2}), and for $v_{r0}$ a value close to $c$, which is predicted by both GR and SRN models (if the energy of the system is high), gives
 \begin{equation} \label{GrindEQ__9}
H_0 = 67.26 \pm 0.90~km/s/Mpc,
\end{equation}
which is in agreement with the observable data.

\begin{figure}[h]
		\includegraphics[width=7.5cm]{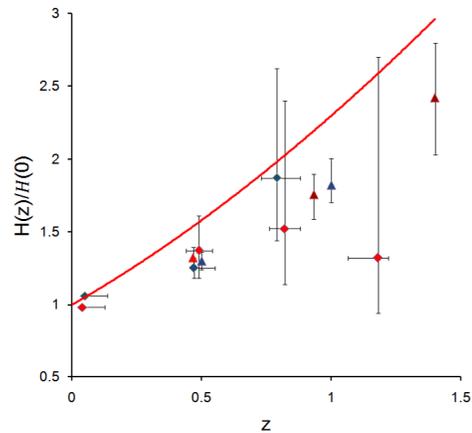}
		\caption{\label{Fig-H_z}
The solid line are calculations for $H(z)/H_0$ at different z using equation (\ref{GrindEQ__10}), with distances corrected for a factor $\pi$, in accordance with our model. Data points are from \cite{4}.
}
\end{figure}

However, let us here rewrite equation (\ref{GrindEQ__8}) in the form
\begin{equation} \label{GrindEQ__10}
                      v_r= H(z)R(z)
\end{equation}
to emphasize that the product $HR$ is equal to the speed of radial expansion and that for models with high energy it is a constant close to $c$. From equation (\ref{GrindEQ__10}) one can also see that the cosmic expansion rate $H(z)$ changed with time and that it must have been larger for an earlier universe. If the universe expanded at an approximately constant speed equation (\ref{GrindEQ__10}) gives for $H(0.5)/H_0 = 1.6$, $H(1.0)/H_0 = 2.3$, and $H(1.4)/H_0 = 3.0$ which agrees with \cite{4} within three standard deviations, Fig. \ref{Fig-H_z}.
Equation (\ref{GrindEQ__10}) is basically the same as the first Friedmann equation
\begin{equation} \label{GrindEQ__11}
                      H^2= \left(\frac{\dot{a}}{a} \right)^2
\end{equation}
where for a closed 3-sphere universe the scale factor $a$ corresponds to the radius of curvature of the universe.  As shown above, our model predicts proper value for $H(z)$. It is important to note that both the Hubble law and the first Friedmann equation directly follow from the model.

One can use GR to make arguments that will favor spherical shell model over $ \Lambda CDM$, which defines the observable universe as a sphere, centered on the observer with the radius $R_0$ = 14.0 $\pm$ 0.2 Gpc (about 45.7 Gly).  To be consistent with general relativity we should require that the model must satisfy GR assumption that the presence of matter or energy causes warping or curvature of spacetime.
The current definition is not consistent with this requirement, because it assumes that light is expanding straight, radially, in all directions for 14 Gpc.

As it is well known if a photon passes a massive object at an impact parameter $b$, the local curvature of space-time will cause the photon to be deflected by an angle
 \begin{equation} \label{mass__1}
 \alpha = \frac{4GM}{c^2b}.
\end{equation}
If photons are in a region of space where gravity is sufficiently strong, a {\it photon sphere} of radius
\begin{equation} \label{photon__1}
 R_{ps} = \frac{3GM}{c^2},
\end{equation}
then the photons will be forced to travel in orbits. It is usually stated that the photon spheres can only exist in the space surrounding an extremely compact object, such as a black hole or a neutron star. However, as it will be shown the concept is also applicable to the visible universe.

The gravitational field {$\bf\Phi$} on the boundary of the imaginary sphere that surrounds mass $M$ is exactly the same as it would have been if all the mass had been concentrated at the center of the sphere.

 By applying Birkhoff's theorem we can assume that the entire mass of the visible universe (considering the visible unverse as a sphere or spherical shell) is located in the center of that sphere. Using for $M$ = $10^{23}M_{\odot}$ = 2x$10^{53}$ kg, it gives $R_{ps}$ = 14.3 Gpc.

 All photons that are inside sphere of 14.3 Gpc must follow circular orbits. Since the visible universe is smaller or about the calculated $R_{ps}$ value all photons will be affected. Therefore, when we speak about the size and the radius of the visible universes we must take into account the bending of light. We cannot say that the visible universe is a sphere with a radius of 14 Gpc since photons cannot travel straight. The measured horizon distance of 14 Gpc is not the length of a straight line, because of the bending of light, it is an arc of a circle with a length of 14 Gpc. The radius of that circle is $14.0/\pi=4.46$ Gpc, which suggest the proposed shell model.

 Because the current observations are not enough accurate to directly test curvature of the space, can we in addition to the above example, which demonstrates that the current model is not taking into account bending of the light predicted by GR, find another test related to the curvature of space.

We will show that holographic principle also favors the shell model over the current $\Lambda$CDM. The entropy of the visible universe is calculated in \cite{3} and it is shown that the dimensionless entropy $S/k$ is 8.85 $\pm$ 0.37 times larger than allowed by a simplified and non-covariant version of the holographic principle, which requires that the entropy cannot exceed that of a black hole.

It was argued in \cite{3} that by the holographic principle the entropy $S/k$ has an upper limit equal to that of a black hole:
\begin{equation} \label{GrindEQ__1}
\left(\frac{S}{k} \right)_{Uni} \le \left(\frac{S}{k} \right)_{BH} =\frac{4\pi R_{S}^{2} }{l_{P}^{2} },
\end{equation}
where $\left(\frac{S}{k} \right)_{Uni} $ is the entropy of the visible universe, $\left(\frac{S}{k} \right)_{BH} $ is the entropy of a black hole, $l_{P}^{} $ is the Planck length, and $R_{S}^{} $ is the Schwarzschild radius $R_{S}^{}$ = 2 $GM$.

Equation (\ref{GrindEQ__1}) requires
\begin{equation} \label{GrindEQ__2}
\frac{\left(\frac{S}{k} \right)_{Uni} }{\left(\frac{S}{k} \right)_{BH} } =R_{BET}^{4} \le 1,
\end{equation}
where $R_{BET}$ is the Bond, Efstathiou, and Tegmark dimensionless shift parameter \cite{3a} defined as
\begin{equation} \label{GrindEQ__3}
                R_{BET} =\frac{\sqrt{\Omega _{m} H_{0}^{2} } }{c} R_0.
\end{equation}
Taking from \cite{2} the size for the radius of the visible universe as $R_0$ = 14.0 $\pm$ 0.2 Gpc gives the value $R_{BET}$= 1.725 $\pm$ 0.018, and hence
\begin{equation} \label{GrindEQ__4}
                             R_{BET}^{4} = 8.85 \pm 0.1,
\end{equation}
which, as pointed out in \cite{3}, is in contradiction with equation (\ref{GrindEQ__2}) for 21$\sigma$. Therefore the current (larger) model for the visible universe  violates the holographic principle. However, the author of \cite{3} at this point speculates that equation (\ref{GrindEQ__2}) was fulfilled in the past when the radius of the universe was
\begin{equation} \label{GrindEQ__5}
                                  R \le 8.4 \pm  0.1~Gpc
\end{equation}
and further speculates that is when the cosmic deceleration ended and  acceleration began.

Both of these assumptions are based on the evaluation of the relation (\ref{GrindEQ__3}) by using the present time Hubble constant $H_0$ instead of the cosmic expansion rate $H(z)$ that corresponds to the size of the universe at that time.
Therefore the radius of the visible universe at past times cannot be obtained from relation (\ref{GrindEQ__3}) by using $H_0$.

According to our shell model $R_0 = 4.46 \pm 0.06~Gpc$
and the ratio in equation (\ref{GrindEQ__2}) is satisfied and it has been always $\le 1$ as it is required by the holographic principle.  Equation (\ref{GrindEQ__3}) is actually another expression for the equation (\ref{GrindEQ__10}), which can be seen by putting in equation (\ref{GrindEQ__3}), $\Omega_m$=1. Therefore, because equation (\ref{GrindEQ__3}) and (\ref{GrindEQ__10}) are the same equations, and because in equation (\ref{GrindEQ__10}) speed $v_{r} \le c$, the inequality (\ref{GrindEQ__2}) must be always $\le 1$ as it is required by the holographic principle.

The presented model has significant consequences for current cosmological theories. It explains uniformity in the CMB without inflation theory. The model also removes any requirements for superluminal speed expansion, since it can explain the size of the universe with speeds less than or equal to the speed of light. However, the size of the observable universe or the radius of the particle horizon is $\pi ct_0$.
 This is between the values $3cH_0^{-1}$ = $3ct_0$, which corresponds to the particle horizon in Einstein-de Sitter universe ($\Omega_m$ = 1 and $\Omega_\Lambda $ = 0) and $3.4ct_0$, which corresponds to the currently favored cosmological model $\Omega_m$ = 0.3 and $\Omega_\Lambda $ = 0.7  \cite{5}. The velocity of the particle horizon of this model is $2c$ which is the same as in the Einstein-de Sitter model.

Our spherical shell model can be further tested. Assuming that matter is homogenously distributed in the universe, a simple experiment which will count the number of galaxies as function of redshift could provide a test for space curvature. If space is in form of a shell, the number of galaxies as function of altitude on the sphere, or function of redshift, should first increase and then decrease. This test is more complex than it appears, since it should take into account the expansion of the space with time and the detection limits of current instrumentation, but it is feasible at the present time. The Hubble Ultra Deep Field (HUDF) optical and near-infrared survey performed in 2004 covered only a tiny patch of the sky, just 3.5 arc minutes across, but due to the high sensitivity and long exposure time extends thousands of megaparsecs away. HUDF shows a uniform distribution of matter by distance. This is consistent with the model, since integration by longitude could result in different number of galaxies for different redshifts, but a survey that will confirm this needs to be performed.

 It is important to note that a hollow shell model completely reproduces the distribution of the entire observed radio sources count for the flux density $S$ from $S\approx 10$ $\mu$Jy to $S\approx 10$ Jy \cite{Condon}.

\section{Conclusion}

 Considered is a model that interprets the visible universe as a spherical shell with radius 4.46 $\pm$ 0.06 Gpc and event horizon as the maximal length of the arc of the shell, which has the size of 14.0 $\pm$ 0.2 Gpc. Consistent with this model, the motion of the light and all galaxies is confined to the volume of the shell, which is radially expanding with the current speed $v_{ro}$ close to $c$. The model predicts Hubble constant $H_0$ = 67.26 $\pm$ 0.90 km/s/Mpc and values for the cosmic expansion rates $H(z)$ that are in agreement with observations. It explains uniformity of the CMB without the inflation theory, because by the model the entire observed CMB originates from a single antipodal point and  for that reason the measured CMB must be exactly the same for all directions of observations, if corrected for Sachs-Wolfe fluctuations caused by large scale structures.  The model predicts correct values for the particle horizon $\pi ct_0$ and the velocity of the particle horizon $2c$.
 Justification for the shell model by GR is bending of the space by mass, which requires that the light trajectory  should be an arc of circle rather than a straight line, if the mass of the universe is taken into account. The model is also favored by holographic principle and it allows to eliminate the established discrepancy between non-covariant version of the holographic principle and the calculated dimensionless entropy, $S/k$, for the visible universe that exceeds the entropy of a black hole, which is due to misinterpretation of the size of the visible universe. The model is in agreement with the distribution of radio sources in space, type SN Ia data, and with HUDF optical and near-infrared survey performed in 2004.

\begin{acknowledgments}
I would like to thank S. Matinyan and I. Filikhin for useful discussions. This work is supported by NSF award HRD-0833184 and NASA grant NNX09AV07A.
\end{acknowledgments}

\end{document}